\documentclass[aps,prl,reprint,superscriptaddress]{revtex4-1}  
\usepackage{amsmath}
\usepackage{empheq}
\usepackage{braket}
\usepackage{amssymb}
\usepackage{times,txfonts}
\usepackage{graphicx}
\usepackage{isotope}
\usepackage{rotating} 
\usepackage{color}
\usepackage{siunitx}
\usepackage{hyperref}

\newcommand{\sSD}{\ket{\Phi}}  
\newcommand{\mSD}{\ket{\Psi}}  
\newcommand{\mSDbra}{\bra{\Psi}}  

\newcommand{\defleft}{\,\raisebox{.176mm}{:}\!\! =}

\newcommand{\pDM}[2]{\gamma^{#1}_{#2}}  
\newcommand{\aOp}[2]{a^{#1}_{#2}} 
\newcommand{\aaOp}[2]{\tilde{a}^{#1}_{#2}} 

\newcommand{\Ias}{\mathbb{A}}           

\newcommand{\intmN}[1]{V_{#1}}
\newcommand{\intNNN}{\intmN{3\mathrm{N}}}                   
\newcommand{\matNNN}[2]{v^{#1}_{#2}}         
\newcommand{\intNNNinNOnB}[1]{\intNNN^{\text{MR-NO}#1\text{B}}}    

\newcommand{\matNO}[2]{w^{#1}_{#2}}         
\newcommand{\matNOtwoB}[2]{\bar v^{#1}_{#2}}         

\newcommand{\Nmaxref}{N_{\max}^{\text{ref}}}
\newcommand{\SRGvalueUnit}[1]{\SI{#1}{\femto\meter\tothe 4}}
\newcommand{\hbarOmega}{\hbar\Omega} 

\newcommand{\Hefour}{\isotope[4]{He}}
\newcommand{\Osixteen}{\isotope[16]{O}}
\newcommand{\Bten}{\isotope[10]{B}}
\newcommand{\Ctwelve}{\isotope[12]{C}}
\newcommand{\Lisix}{\isotope[6]{Li}}

\newcommand{\Nmax}{N_{\text{max}}}          

\newcommand{\SRGpara}{\alpha}

\begin{document}


\title{Open-Shell Nuclei and Excited States from Multi-Reference Normal-Ordered Hamiltonians}
 
\author{Eskendr Gebrerufael}
\email[]{eskendr@theorie.ikp.physik.tu-darmstadt.de}
\affiliation{Institut f\"ur Kernphysik, TU Darmstadt, Schlossgartenstr.\ 2, 64289 Darmstadt, Germany}

\author{Angelo Calci}
\affiliation{TRIUMF, 4004 Wesbrook Mall, Vancouver, British Columbia, V6T 2A3 Canada}

\author{Robert Roth}
\affiliation{Institut f\"ur Kernphysik, TU Darmstadt, Schlossgartenstr.\ 2, 64289 Darmstadt, Germany}

\date{\today}

\begin{abstract}

We discuss the approximate inclusion of three-nucleon interactions into \emph{ab initio} nuclear structure calculations using a multi-reference formulation of normal ordering and Wick's theorem. Following the successful application of single-reference normal ordering for the study of ground states of closed-shell nuclei, e.g., in coupled-cluster theory, multi-reference normal ordering opens a path to open-shell nuclei and excited states. Based on different multi-determinantal reference states we benchmark the truncation of the normal-ordered Hamiltonian at the two-body level in no-core shell-model calculations for p-shell nuclei, including $\Lisix$, $\Ctwelve$ and $\Bten$. We find that this multi-reference normal-ordered two-body approximation is able to capture the effects of the 3N interaction with sufficient accuracy, both, for ground-state and excitation energies, at the computational cost of a two-body Hamiltonian. It is robust with respect to the choice of reference states and has a multitude of applications in \emph{ab initio} nuclear structure calculations of open-shell nuclei and their excitations as well as in nuclear reaction studies.

\end{abstract}

\pacs{21.60.De, 21.30.-x, 05.10.Cc, 21.45.Ff}

\maketitle


\paragraph{Introduction.}

Over the past few years one of the major advances for many \emph{ab initio} many-body approaches, particularly for medium-mass nuclei, was the inclusion of three-nucleon (3N) interactions. They arise naturally in nuclear interactions constructed in chiral effective field theory \cite{RevModPhys.81.1773,Machleidt20111,Weinberg1990288} and are inevitable when working with softened Hamiltonians obtained, e.g., from a similarity renormalization group (SRG) transformation \cite{PhysRevLett.107.072501,PhysRevC.75.061001,PhysRevLett.103.082501,PhysRevC.90.024325}. Generally, a systematic and consistent inclusion of 3N interactions are a prime goal of modern nuclear structure and reaction theory. 

Compared to many-body calculations with two-nucleon (NN) forces only, the inclusion of 3N interactions leads to a significant increase of the computational and formal complexity of the many-body problem. In some cases, like the no-core shell model (NCSM) \cite{NaGuNo07}, the formal inclusion is straight-forward, but the increase of the computational cost is significant \cite{Vary2009}. In other cases, like coupled-cluster (CC) theory for nuclear ground states \cite{Binder2014119,PhysRevC.87.021303}, the complexity of the basic many-body equations and of the numerical solution both increase dramatically. Similarly, in calculations of continuum and scattering observables, e.g., in the NCSM combined with the resonating-group method (NCSM/RGM) \cite{PhysRevLett.101.092501,PhysRevC.79.044606,PhysRevC.88.054622} or the NCSM with continuum (NCSMC) \cite{PhysRevLett.110.022505,PhysRevC.87.034326}, the inclusion of explicit 3N interactions is feasible only in simple cases \cite{PhysRevC.90.061601,PhysRevC.91.021301}.

Therefore, approximation schemes are highly desirable that include the physics of 3N interactions at the cost of a calculation with only NN interactions. To this end, effective or phenomenological NN interactions that are adjusted to capture some physics aspects of the 3N force have been constructed in the past. A more systematic way to derive such approximations starts from the normal-ordered form of the Hamiltonian with respect to an $A$-body reference state. 

Normal ordering of products of creation and annihilation operators with respect to non-trivial reference states is an important technical element in the formulation of a number of modern many-body approaches. Two of the most successful \emph{ab initio} approaches for medium-mass nuclei, CC theory \cite{Coester1958421,PhysRevLett.94.212501,PhysRevLett.101.092502,HaPa07-CC} and the in-medium similarity renormalization group (IM-SRG) \cite{PhysRevLett.106.222502,PhysRevC.85.061304}, are constructed in a normal-ordered formulation from the outset. In addition to the formal advantages of working with normal-ordered products, normal-ordering also presents a natural starting point for the approximate inclusion of multi-nucleon interactions. Already the normal-ordered zero-, one- and two-body terms of the Hamiltonian contain contributions of the initial 3N interaction. Thus, by truncating the Hamiltonian beyond the normal-ordered two-body level, we take into account parts of the 3N interaction, while retaining the computational complexity of a calculation with only two-body terms. 

This normal-ordered two-body (NO2B) approximation has been successfully applied in recent many-body calculations, particularly in the medium-mass regime \cite{PhysRevC.90.041302,Binder2014119,PhysRevLett.113.142502,PhysRevC.87.011303,PhysRevLett.111.062501}. We have studied the quality of the NO2B approximation in NCSM and CC calculations for ground states of closed-shell nuclei by direct comparison to calculations with explicit 3N interactions within the same many-body framework. Within the NCSM we have tested the NO2B approximation for \Hefour\ and \Osixteen\ and found deviations from the ground-state energies obtained with explicit 3N interaction on the order of 2\% and 1\%, respectively \cite{Roth:2011vt}. We also extended CC theory to include explicit 3N interaction at the singles and doubles level \cite{PhysRevC.87.021303,HaPa07-CC} and with non-iterative triples corrections \cite{PhysRevC.88.054319}. This enabled a direct benchmark of the NO2B approximation in the medium-mass regime, which robustly confirmed that this approximation agrees with ground-state results with the explicit 3N interactions to better than 1\%. Given the other uncertainties in medium-mass approaches, this is acceptable for many applications. 

In this work, we generalize and test the NO2B approximation to ground-state and excitation energies of open-shell nuclei through a generalization of normal ordering to multi-determinantal reference states. After discussing the formalism we will benchmark the multi-reference normal-ordered two-body (MR-NO2B) approximation in NCSM calculations for the ground and excited states of p-shell nuclei, comparing directly to calculations with explicit 3N interactions.

\paragraph{Normal-Ordered Hamiltonian.}

In the simplest formulation, a product of creation and annihilation operators is normal ordered if the creators are to the left of all annihilators. In the following we use a convenient short-hand notation for creation operators $\aOp{p}{} \defleft a_p^\dagger$ and annihilation operators $\aOp{}{p}$ defined with respect to a complete orthonormal single-particle basis $\{ \ket{p} \}$. Furthermore, we write the particle-hole operators   
\begin{equation}\begin{split}
  \label{eq:phoperators}
  \aOp{prt \ldots }{qsu \ldots} &\defleft \aOp{p}{} \aOp{r}{} \aOp{t}{} \cdots \aOp{}{u} \aOp{}{s} \aOp{}{q} 
\end{split}\end{equation}
for single-particle indices covering the full single-particle basis. These particle-hole operators are naturally in normal order with respect to the vacuum state $\ket{0}$ and the expectation value of these normal-ordered products with the vacuum state vanishes. This is a necessary criterion for normal-ordered products with respect to a specific reference state. This vacuum normal-order corresponds to the standard representation of second quantized operators, as they are used, e.g., in NCSM calculations. Specifically for the operator of a 3N interaction, the vacuum normal-ordered form is given by 
\begin{equation}
  \label{eq:v3n}
  \intNNN =  \frac{1}{36}  \sum_{\substack{prt \\ qsu}} \matNNN{prt}{qsu}\; \aOp{prt}{qsu}
\end{equation}
with antisymmetrized three-body matrix elements $\matNNN{prt}{qsu}\nobreakspace\defleft\nobreakspace\bra{prt}\intNNN\ket{qsu}$.

The more interesting case, however, is normal ordering with respect to an $A$-body reference state. Assuming the simple case of an $A$-body reference state $\sSD$ given by a single Slater determinant built from $A$ occupied single-particle states, this leads to the standard picture of particle-hole excitations on top of the reference state. Particle states are unoccupied in the reference states and hole states are occupied. The notion of normal ordering needs to be extended to guarantee that expectation values of any string of normal-ordered operators in the reference state vanishes. To this end, particle creation and hole annihilation operators are to the left of particle annihilation and hole creation operators, defining the single-reference normal ordering. 

One can convert from vacuum to single-reference normal order by explicit use of the anticommutation relations for fermionic creation and annihilation operators or, more elegantly, through Wick's theorem \cite{PhysRev.80.268}. We have formulated and benchmarked the use of the single-reference normal ordering and the NO2B approximation for closed-shell nuclei in detail in Ref.~\cite{Roth:2011vt} using NCSM calculation and in Refs.~\cite{PhysRevC.87.021303,PhysRevC.88.054319} using the CC framework. 

The single-reference formulation of normal ordering is linked directly to the notion of particle and hole states. For the generalization to more complicated reference states $\mSD$, given by a superposition of many Slater determinants, a distinction of particle and hole states is not possible anymore. Therefore, normal ordering for multi-determinantal reference states operates on a more formal level and we rely entirely on generalizations of Wick's theorem.  

We adopt the multi-reference version of the Wick's theorem proposed and proven by Kutzelnigg and Mukherjee~\cite{kutzelnigg:432, Mukherjee1997561}. The nontrivial contractions correspond to the irreducible $n$\nobreakdash-body density matrix elements encoding information about $n$\nobreakdash-body correlations in the reference state, which can be expressed in terms of the $m$\nobreakdash-body density matrix elements
\begin{equation}\begin{split}
  \pDM{p_1 p_2 \ldots p_m}{q_1 q_2 \ldots q_m} &= \mSDbra \aOp{p_1 p_2 \ldots p_m}{q_1 q_2 \ldots q_m}  \mSD
\end{split}\end{equation}
with $m\leq n$. Applying the multi-reference Wick's theorem to rewrite the particle-hole operators $\aOp{p}{q}$, $\aOp{pr}{qs}$, $\aOp{prt}{qsu}$ of Eq.~\eqref{eq:phoperators} in terms of multi-reference normal-ordered particle-hole operators $\aaOp{p}{q}$, $\aaOp{pr}{qs}$, $\aaOp{prt}{qsu}$, we obtain after simplifications~\cite{kutzelnigg:432}
\begin{equation}\begin{aligned}
  \label{eq:phoperators-no}
    \aOp{p}{q} &= \aaOp{p}{q} +  \pDM{p}{q} \;, \\
	  \aOp{pr}{qs} &= \aaOp{pr}{qs} + \Ias( \pDM{p}{q} \aaOp{r}{s} ) + \pDM{pr}{qs} \;, \\
	  \aOp{prt}{qsu} &= \aaOp{prt}{qsu} + \Ias( \pDM{p}{q} \aaOp{rt}{su} ) +  \Ias( \pDM{pr}{qs} \aaOp{t}{u} ) + \pDM{prt}{qsu} \;,
\end{aligned}\end{equation}
where $\Ias$ is the index antisymmetrizer as defined in Ref.~\cite{kong:234107} generating a totally antisymmetric sum of all possible permutations within the upper and lower indices avoiding duplicates.
 
Inserting \eqref{eq:phoperators-no} into the second-quantized form of the 3N interaction \eqref{eq:v3n}, we obtain the 3N interaction in multi-reference normal-ordered form
\begin{gather}
  \label{eq:v3n-no}
  \intNNN = \matNO{}{} + \sum_{\substack{p \\ q}} \matNO{p}{q}\, \aaOp{p}{q} + \frac{1}{4} \sum_{\substack{pr \\ qs}} \matNO{pr}{qs}\, \aaOp{pr}{qs} + \frac{1}{36}  \sum_{\substack{prt \\ qsu}} \matNO{prt}{qsu}\, \aaOp{prt}{qsu}
  \intertext{with}
  \begin{aligned}
  \matNO{}{} &= \frac{1}{36} \sum_{\substack{prt \\ qsu}} \matNNN{prt}{qsu} \pDM{prt}{qsu} \;, &
  \matNO{p}{q} &= \frac{1}{4} \sum_{\substack{rt \\ su}} \matNNN{prt}{qsu} \pDM{rt}{su} \;, \\
  \matNO{pr}{qs} &= \sum_{\substack{t \\ u}} \matNNN{prt}{qsu} \pDM{t}{u} \;, &
  \matNO{prt}{qsu} &= \matNNN{prt}{qsu}\;.
  \end{aligned}
\end{gather}
The matrix elements of the multi-reference normal-ordered $n$-body contributions emerging from the 3N interaction are given by simple summations of the original three-body matrix elements contracted with density matrices. Inserting the density matrices obtained for a single-determinant reference state immediately reduces these expressions to the known single-reference expressions \cite{Roth:2011vt,PhysRevC.87.021303}.  

By omitting the normal-ordered three-body contribution in \eqref{eq:v3n-no} we define the MR-NO2B approximation of the 3N interaction
\begin{equation}
  \label{eq:v3n-no2b-vac}
  \intNNNinNOnB{2} 
  = \matNO{}{} + \sum_{\substack{p \\ q}} \matNO{p}{q}\, \aaOp{p}{q} + \frac{1}{4} \sum_{\substack{pr \\ qs}} \matNO{pr}{qs}\, \aaOp{pr}{qs} \; .
\end{equation}
For approaches like the NCSM that do not naturally use a normal-ordered formulation, we can invert the relations \eqref{eq:phoperators-no} to convert the above expression back to vacuum normal order and obtain  
\begin{equation}
  \label{eq: 3N interaction in MR-NO2B approximation denoted in vacuum representation}
  \intNNNinNOnB{2}  
   = \matNOtwoB{}{} +  \sum_{\substack{p \\ q}} \matNOtwoB{p}{q}\;  \aOp{p}{q} + \frac{1}{4} \sum_{\substack{pr \\ qs}} \matNOtwoB{pr}{qs}\;  \aOp{pr}{qs}
\end{equation}
with
\begin{equation}\begin{aligned}
  \label{eq:v3n-no2b-vac-me}
  \matNOtwoB{}{} &= \frac{1}{36} \sum_{\substack{prt \\ qsu}} \matNNN{prt}{qsu} \left(\pDM{prt}{qsu} - 18 \pDM{p}{q} \pDM{rt}{su} + 36 \pDM{p}{q} \pDM{r}{s} \pDM{t}{u} \right) \;,\\ 
  \matNOtwoB{p}{q} &= \frac{1}{4} \sum_{\substack{rt \\ su}} \matNNN{prt}{qsu} \big(\pDM{rt}{su} - 4\pDM{r}{s} \pDM{t}{u}\big) \;,\qquad  
  \matNOtwoB{pr}{qs} = \sum_{\substack{t \\ u}} \matNNN{prt}{qsu} \pDM{t}{u} \;.
\end{aligned}\end{equation}
We remark that we do not need the three-body density matrix element explicitly since the term including the three-body density matrix element in the zero-body part can be identified as the expectation value of the 3N interaction in the reference state, which can be computed directly. The one- and two-body density matrices can be easily computed using standard NCSM technology.

\paragraph{Calculation Details.}
The starting point of our calculations is a Hamiltonian based on the NN or NN+3N interaction from chiral effective field theory. We use the NN interaction at next-to-next-to-next-to-leading ($\text{N}^3\text{LO}$) from Entem and Machleidt~\cite{EnMa2003PhysRevC.68.041001} and the 3N interaction at $\text{N}^3\text{LO}$ in local form from Navr\'atil~\cite{Na2007FewBodySyst}. The low-energy constants have been fitted to the ground-state energy  and $\beta$-decay half-life of 3N systems~\cite{GaDoQu2009PhysRevLett.103.102502}.
Both Hamiltonians will be transformed by means of the SRG in three-body space, in order to enhance convergence behavior with respect to the many-body model space \cite{PhysRevLett.107.072501,PhysRevC.77.064004}. Here, we consider two types of SRG-evolved Hamiltonians:
The NN+3N-induced Hamiltonian omits the chiral 3N interaction from the initial Hamiltonian, but keeps all induced 3N terms throughout the transformation;
the NN+3N-full Hamiltonian starts with the initial chiral NN+3N Hamiltonian and retains all terms up to the three-body level in the SRG transformation. The 3N-interaction terms in both Hamiltonians have quite different characteristics, which makes them useful for benchmarking the MR-NO2B approximation. 

For each of these Hamiltonians, we apply the MR-NO2B approximation with respect to nucleus-specific reference states. These reference states are given by the ground state obtained from full NCSM calculations in small model spaces, characterized by $\Nmaxref$, including explicit three-body interactions. In order to analyze the dependence on the reference state, we vary the truncation parameter $\Nmaxref$ and, thus, obtain a sequence of reference states and a corresponding sequence of MR-NO2B approximations. Note that for closed-shell nuclei the MR-NO2B approximation with $\Nmaxref=0$ is equivalent to the single-reference version of the NO2B approximation. 

Finally, we use the MR-NO2B matrix elements in importance-truncated no-core shell model (IT-NCSM) calculations for ground and excited states of p-shell nuclei up to large model-space truncations $N_{\max}$. To remove spurious center-of-mass excitations from the low-energy spectra we add a harmonic oscillator center-of-mass Hamiltonian. In order to benchmark the MR-NO2B approximation we compute the same observables in the IT-NCSM including explicit 3N interactions. These calculations are significantly more expensive, since the explicit three-body terms reduce the sparsity of the many-body Hamilton matrix, i.e., increase the number of non-zero matrix elements by at least an order of magnitude \cite{Vary2009}. Details on the IT-NCSM can be found in Refs.~\cite{PhysRevC.79.064324,PhysRevLett.99.092501}.

\begin{figure}
	\includegraphics[width=1\columnwidth]{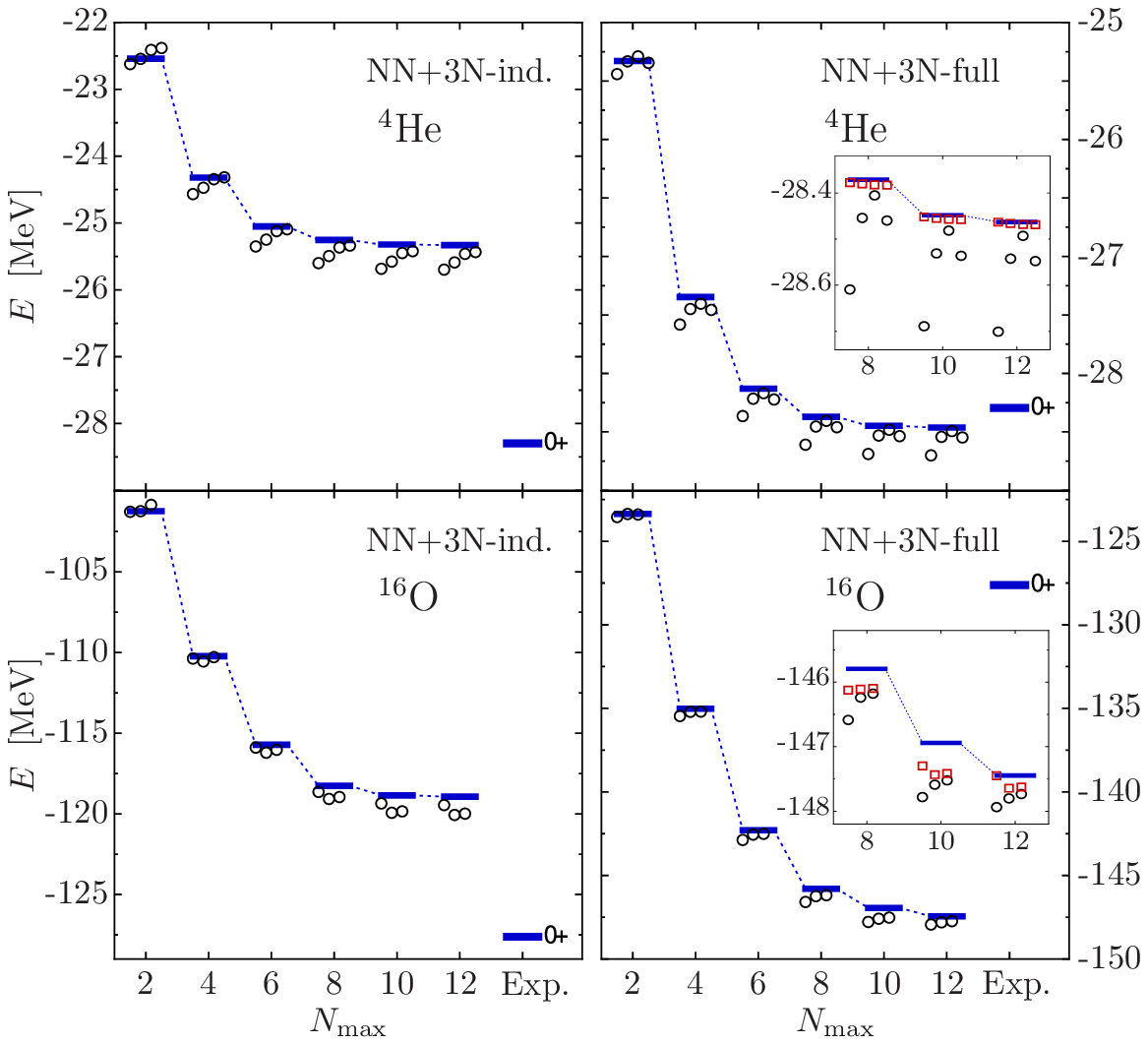}
	\caption{(color online) IT-NCSM absolute ground-state energies of $\Hefour$ (top) and $\Osixteen$ (bottom) as function of $\Nmax$ for the NN+3N-induced (left) and NN+3N-full (right) Hamiltonian with the SRG flow parameter $\SRGpara=\SRGvalueUnit{0.08}$ and $\hbarOmega=\SI{20}{\MeV}$. The levels connected by a dashed line correspond to the complete 3N interaction, the circles to MR-NO2B approximations for a range of $\Nmaxref$ parameters: $\Nmaxref$ = 0, 2, 4, 6 (from left to right). The insets for the NN+3N-full calculations show results with a perturbative inclusion of the residual normal-ordered 3N term as red boxes (see text). Experimental ground-state energies are taken from \cite{WaAuWa12}. }
	\label{fig:He4 and O16}
\end{figure}

\paragraph{Closed-Shell Nuclei.}
We start with a direct comparison of IT-NCSM ground-state energies for the closed-shell nuclei $\Hefour$ and $\Osixteen$. These nuclei have been already investigated in the framework of the single-reference NO2B approximation~\cite{Roth:2011vt}. The multi-reference formulation can go beyond this single-determinant reference and explore the impact of improved reference states for larger $\Nmaxref$, which systematically approach the converged ground state of the nucleus.

In Fig.~\ref{fig:He4 and O16} we present the absolute ground-state energies of $\Hefour$ and $\Osixteen$ as function of $\Nmax$ calculated for the NN+3N-induced and NN+3N-full Hamiltonian with the SRG flow parameter $\SRGpara=\SRGvalueUnit{0.08}$. All MR-NO2B results are in good agreement with the calculations including explicit 3N terms, the largest relative deviations are at the level of 1\%.

Closer inspection of the results with increasing $\Nmaxref$ shows that there is no universal systematics. For $\Hefour$ the agreement with the full calculation improves when going from $\Nmaxref=0$ to $2$, but for $\Osixteen$ the agreement gets slightly worse in case of the NN+3N-induced Hamiltonian. Generally the dependence on the reference state, i.e. on $\Nmaxref$ is small, indicating that the MR-NO2B approximation is robust with respect to variations of the reference state. 
To improve on the MR-NO2B approximation, we can attempt to include the residual normal-ordered three-body terms perturbatively by adding their expectation value obtained with the MR-NO2B eigenstates. The results for selected cases are shown in the insets in Fig.~\ref{fig:He4 and O16}. This correction does improve the agreement with the full calculation, but generally cannot remove the difference completely, as evident from the $\Osixteen$ results. The remaining different has to be attributed to differences in the MR-NO2B eigenstates compared to the eigenstates of the complete Hamiltonian.
Furthermore, we note that for these and all following cases, the expectation values of total angular momentum and harmonic-oscillator center-of-mass Hamiltonian obtained with the MR-NO2B approximation and with explicit 3N interactions do agree within the numerical accuracy of the IT-NCSM, indicating rotational and translational invariance of the MR-NO2B Hamiltonian.

\begin{figure}
	\centering
	\includegraphics[width=1\columnwidth]{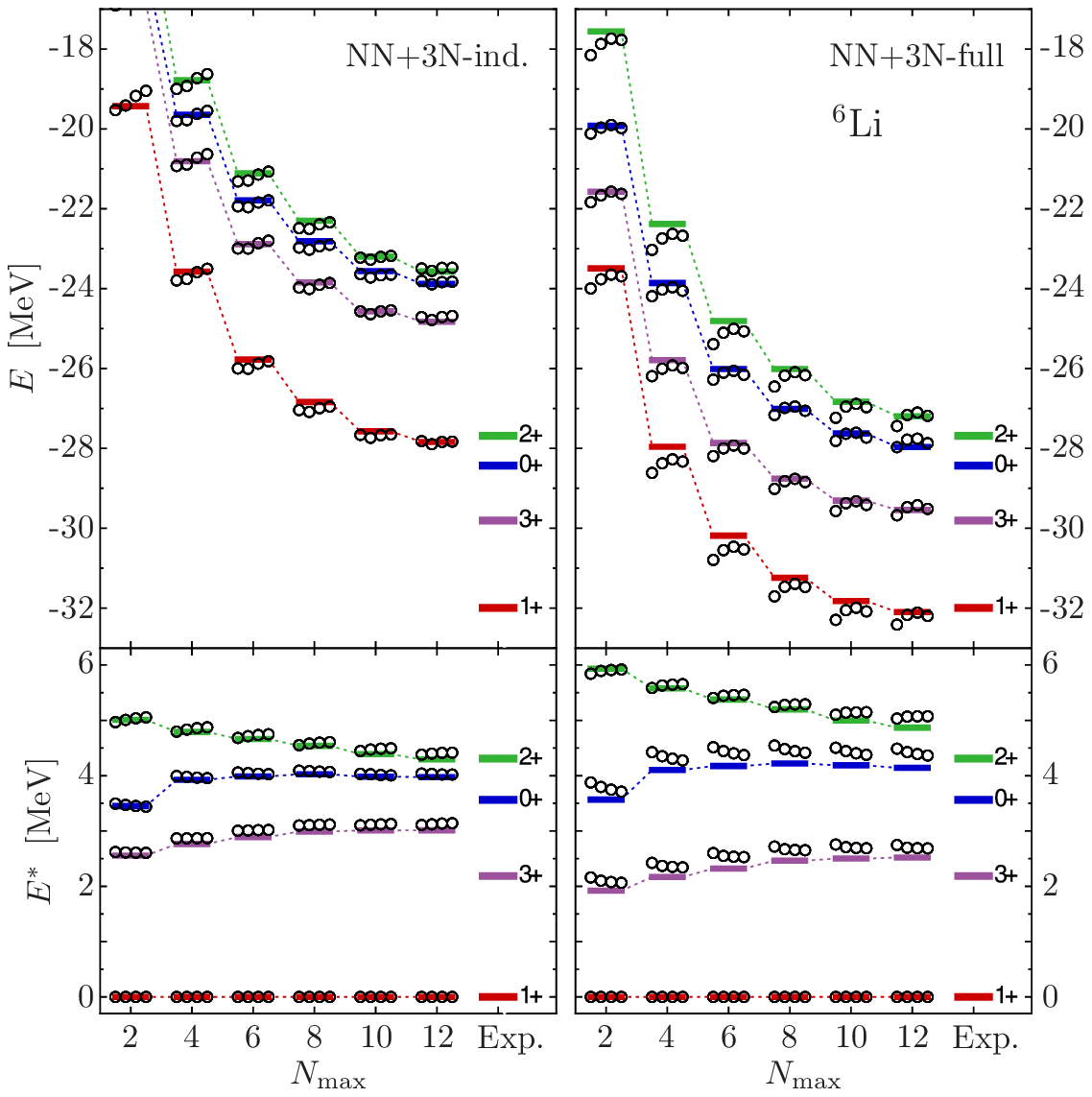}
	\caption{(color online) IT-NCSM absolute (top) and relative (bottom) spectrum of $\Lisix$ as function of $\Nmax$ for the NN+3N-induced (left) and NN+3N-full (right) Hamiltonian with the SRG flow parameter $\SRGpara=\SRGvalueUnit{0.08}$ and $\hbarOmega=\SI{20}{\MeV}$. The solid levels connected by a dashed line correspond to the complete 3N interaction, the circles to MR-NO2B approximations for a range of $\Nmaxref$ parameters: $\Nmaxref$ = 0, 2, 4, 6 (from left to right). Experimental excitation energies are taken from \cite{nndc}.}
	\label{fig:Li6}
\end{figure}
\begin{figure}
 	\centering
	\includegraphics[width=1\columnwidth]{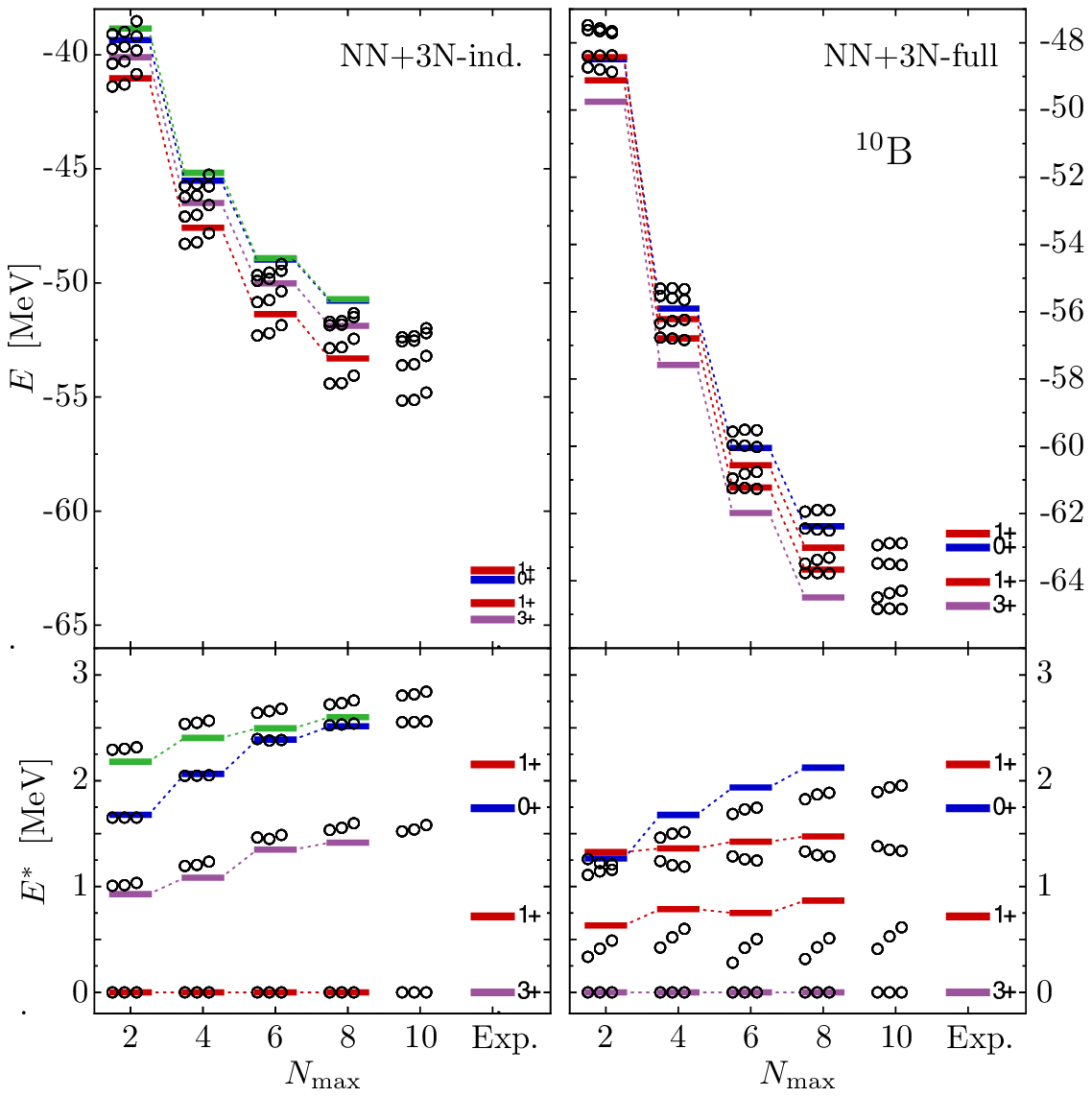}
	\caption{(color online) Same as Fig.~\ref{fig:Li6} for $\Ctwelve$ with $\Nmaxref$ = 0, 2, 4 (from left to right).}
	\label{fig:C12}
\end{figure}

\paragraph{Open-Shell Nuclei.}
The multi-reference formulation now allows us to address open-shell nuclei as well. We will investigate the ground-state and excitation energies of $\Lisix$, $\Ctwelve$ and $\Bten$ as a representative set of p-shell nuclei. The reference state for the MR-NO2B approximation is always the ground state from full NCSM calculations with small $\Nmaxref$, also for the calculation of the excited states. Thus, we will address two aspects: the quality of the  MR-NO2B approximation for the description of the ground-state in open-shell systems and the quality of the normal-ordering based on the ground state as reference for the description of excited states.  

Figure~\ref{fig:Li6} shows the absolute energies of $\Lisix$ for the four lowest natural-parity states as well as the excitation energies as function of $\Nmax$. The calculations are carried out using the IT-NCSM for the NN+3N-induced and NN+3N-full Hamiltonian with the SRG flow parameter $\SRGpara=\SRGvalueUnit{0.08}$. As for the closed-shell cases we use different $\Nmaxref$ to vary the complexity of the reference state and compare to direct IT-NCSM calculations with explicit 3N interactions.

For the absolute energies we observe an excellent agreement of the various levels of the MR-NO2B approximation with the calculations including the 3N interactions explicitly. Particularly, there is no difference in the quality of the description of the ground state and the excited states. This point is important given the fact that the normal ordering is performed for a reference state that is constructed as an approximation for ground state and does not include information about the excited states. 

As function of $\Nmaxref$ the absolute energies of all states show the same systematics and, as a result, the excitations show a smooth and very weak dependence on the reference state. As for the ground states of closed-shell nuclei, we generally do not observe a systematic improvement of the MR-NO2B results with increasing $\Nmaxref$.

\begin{figure}
 	\centering
	\includegraphics[width=1\columnwidth]{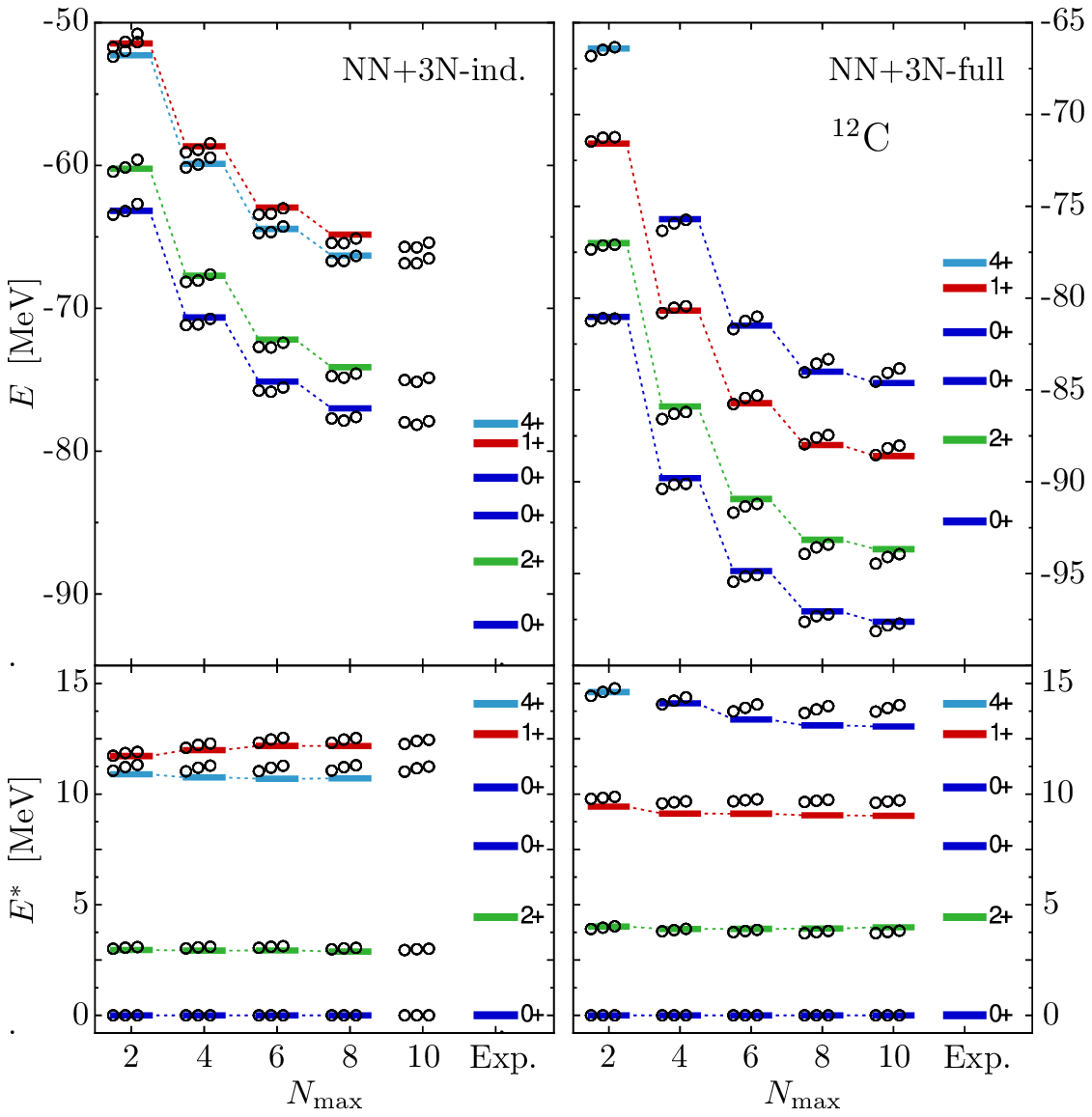}
	\caption{(color online) Same as Fig.~\ref{fig:Li6} for $\Bten$ with $\hbarOmega=\SI{16}{\MeV}$ and $\Nmaxref$ = 0, 2, 4 (from left to right).}
	\label{fig:B10}
\end{figure}

To test the MR-NO2B approximation in nuclei with a more complicated structure, we consider the low-lying natural-parity states in $\Ctwelve$ and $\Bten$ shown in Figs.~\ref{fig:C12} and \ref{fig:B10}, respectively. Previous investigations have shown that several states in these nuclei are sensitive to the chiral 3N interaction \cite{NaGu07,PhysRevLett.107.072501} and, thus, are critical tests for the MR-NO2B approximation. Also, these calculations are more challenging from the point of view of the importance truncation and threshold extrapolations. We have benchmarked the IT-NCSM against full NCSM calculations for $\Ctwelve$ in a previous publication \cite{PhysRevC.90.014314} showing that the uncertainties due to the threshold extrapolation in the IT-NCSM are small on the scales discussed here and that they can be estimated reliably. For the excitation energies, we expect maximum extrapolation uncertainties on the order of $0.1$ MeV for the largest $N_{\max}$ considered here. This is insignificant for $\Ctwelve$ but not completely negligible for $\Bten$.

For $\Ctwelve$ the agreement of the MR-NO2B approximation with the full calculations is at a similar level as for the simpler nucleus $\Lisix$. The absolute energies show the same trends for ground and excited states as function of $\Nmaxref$ and, consequently, the excitation energies show a mild dependence on the reference state. As before, larger $\Nmaxref$ do not necessarily improve the MR-NO2B approximation. Generally, the MR-NO2B approximation works very well and the deviations for the results with explicit 3N interactions are at the same level as the uncertainties due to $N_{\max}$-convergence and threshold extrapolation. 

At the same time, the computational cost for the IT-NCSM calculations with the MR-NO2B approximation is an order of magnitude lower than for the full 3N calculations, because of a significant reduction of the number of non-zero matrix elements in the many-body Hamilton matrix, which entails small importance truncated spaces. For the $N_{\max}=10$ calculations shown in Fig.~\ref{fig:C12} the total CPU time for the IT-NCSM calculation reduces by a factor $10$ and the maximum memory footprint by a factor $20$.   

The most challenging test case is clearly $\Bten$. Being an odd-odd nucleus, the excitation energies are smaller and deviations become more significant. Moreover, it is known that for the present chiral Hamiltonian, the 3N interaction is responsible for changing the ordering of the two lowest states \cite{Barrett2013131}. Using the chiral NN interaction only, the $1^+$ states emerges as ground state in contradiction to experiment and only the inclusion of the chiral 3N interaction leads to the correct $3^+$ ground state. Thus the 3N force has a drastic impact on the spectrum and it is unclear whether the MR-NO2B approximation is able to capture this effect.

The results for $\Bten$ depicted in Fig.~\ref{fig:B10} show that the MR-NO2B approximation can cope with these situations with a somewhat reduced accuracy. The absolute energies obtained in the MR-NO2B approximation deviate by up to $1$ MeV from the full 3N results with uncertainties due to the IT-NCSM threshold extrapolations up to $0.3$ MeV. For the excitation energies the deviations are significantly smaller, particularly for the NN+3N-induced Hamiltonian where the absolute deviations are at the same level as for the simpler nuclei. For the NN+3N-full Hamiltonian, the deviations of the excitation energies reach $0.5$ MeV and show a stronger dependence on $\Nmaxref$. However, the MR-NO2B approximations always give the correct level ordering and, thus, capture the most important effects of the 3N interaction. Moreover, the larger dependence on the reference state and $\Nmaxref$ can serve as an indicator for the reduced quality of the MR-NO2B approximation. 

\paragraph{Conclusions.} 

We have introduced and studied the MR-NO2B approximation for the efficient inclusion of the 3N interactions in nuclear structure calculations for ground and excited states of open-shell nuclei. Through direct comparison with IT-NCSM calculations including the full 3N interactions explicitly we have demonstrated the robustness and accuracy of this approximation. The absolute energies of ground and excited states for closed- and open-shell nuclei typically agree with the full 3N results at the 1\% level, with the exception of very light nuclei (e.g., $\Hefour$) and particularly fragile states (e.g., $\Bten$). The description of excited states exhibits the same quality and systematics as the ground states although the normal ordering only involves a reference state representative for the ground state. This, together with the small dependence on the specific choice of the reference states, i.e., the $\Nmaxref$, demonstrates the robustness of the MR-NO2B approximation for the 3N interaction.

These findings have important implications for a range of many-body applications. In the context of the NCSM and IT-NCSM the MR-NO2B approximation gives access to nuclei that are computationally out of reach with explicit 3N interactions. Due to the significant reduction of the number of non-zero matrix elements in the many-body Hamilton matrix, nuclei in the lower half of the sd-shell become accessible in the IT-NCSM at manageable computational cost. The MR-NO2B approximation for 3N interactions also facilitates continuum and reaction calculations in the NCSM/RGM and NCSMC, which are too demanding with explicit 3N terms. For medium-mass approaches, particularly the IM-SRG \cite{PhysRevLett.110.242501,PhysRevC.87.034307,PhysRevC.90.041302}, which are formulated with normal-ordered operators from the outset, the quality of the MR-NO2B approximation directly affects the accuracy of the whole many-body framework. Our findings for excited states also open the door towards studies of excitation spectra with evolved Hamiltonians from the IM-SRG, e.g., through subsequent equations-of-motion or configuration-interaction calculations. Furthermore, the normal-ordering framework can be directly extended to 4N interactions, making the inclusion of chiral 4N forces that emerge at order N${}^3$LO possible for a variety of many-body approaches. 
   
\begin{acknowledgments}
\paragraph{Acknowledgments.}
We thank H. Hergert for useful discussions. This work is supported by the DFG through contract SFB 634, the Helmholtz International Center for FAIR (HIC for FAIR), the BMBF through contract 05P15RDFN1. TRIUMF receives funding via a contribution through the National Research Council Canada. The authors gratefully acknowledge computing time granted by the CSC Frankfurt (LOEWE-CSC), the J\"ulich Supercomputer Center (JUROPA), and the computing center of the TU Darmstadt (LICHTENBERG).
\end{acknowledgments}
 

%

\end{document}